\documentclass[aps,twocolumn,superscriptaddress,showpacs,showkeys,preprintnumbers,amsmath,amssymb,amsfonts,nofootinbib]{revtex4}

\pdfoutput=1

\usepackage{graphicx}
\usepackage{color}
\usepackage{slashed}
\usepackage{tabularx}
\usepackage{hyperref}

\newcommand{\I}{\mathrm{i}}

\definecolor{bjcol}{rgb}{0.12,0.56,1}

\newcolumntype{L}[1]{>{\raggedright\arraybackslash}p{#1}} 
\newcolumntype{C}[1]{>{\centering\arraybackslash}p{#1}} 
\newcolumntype{R}[1]{>{\raggedleft\arraybackslash}p{#1}} 

\graphicspath{{.}{figures/}{mma/}}

\begin{document}

\title{The low-temperature behavior of the quark-meson model}

\newcommand{\TUD}{Theoriezentrum, Institut f\"ur Kernphysik, Technische Universit\"at Darmstadt, Schlossgartenstrasse 2, 64289 Darmstadt, Germany}
\newcommand{\JLU}{Institut f\"ur Theoretische Physik, Justus-Liebig-Universit\"at Giessen, Heinrich-Buff-Ring 16, 35392 Giessen, Germany}
\newcommand{\ECT}{European Centre for Theoretical Studies in Nuclear Physics and Related Areas (ECT*) and Fondazione Bruno Kessler, Villa Tambosi, Strade delle Tabarelle 286, I-38123 Villazzano (TN), Italy}

\author{Ralf-Arno Tripolt}\affiliation{\ECT}
\author{Bernd-Jochen Schaefer}\affiliation{\JLU}
\author{Lorenz von Smekal}\affiliation{\JLU}
\author{Jochen Wambach}\affiliation{\ECT}

\begin{abstract}
We revisit the phase diagram of strong-interaction matter for the two-flavor quark-meson model using the Functional Renormalization Group. In contrast to standard mean-field calculations, an unusual phase structure is encountered at low temperatures and large quark chemical potentials. In particular, we identify a regime where the pressure decreases with increasing temperature and discuss possible reasons for this unphysical behavior.
\end{abstract}



\maketitle

\section{Introduction}\label{sec:intro}

As an effective low-energy description of Quantum Chromodynamics (QCD) the quark-meson (QM) model is widely used to compute the phase diagram of strong-interaction matter. The model shares the chiral symmetry-breaking pattern of QCD and is renormalizable. When imposing spatial homogeneity a first-order chiral transition at low temperature is found in the mean-field (MF) approximation, which ends in a critical point. In this point the phase transition is of second order and lies in the same universality class as the liquid-gas transition. Such a `chiral critical point' is searched for in dedicated experimental programs such as the beam-energy scan at RHIC. 

\begin{figure}[t]
	\includegraphics[width=0.48\textwidth]{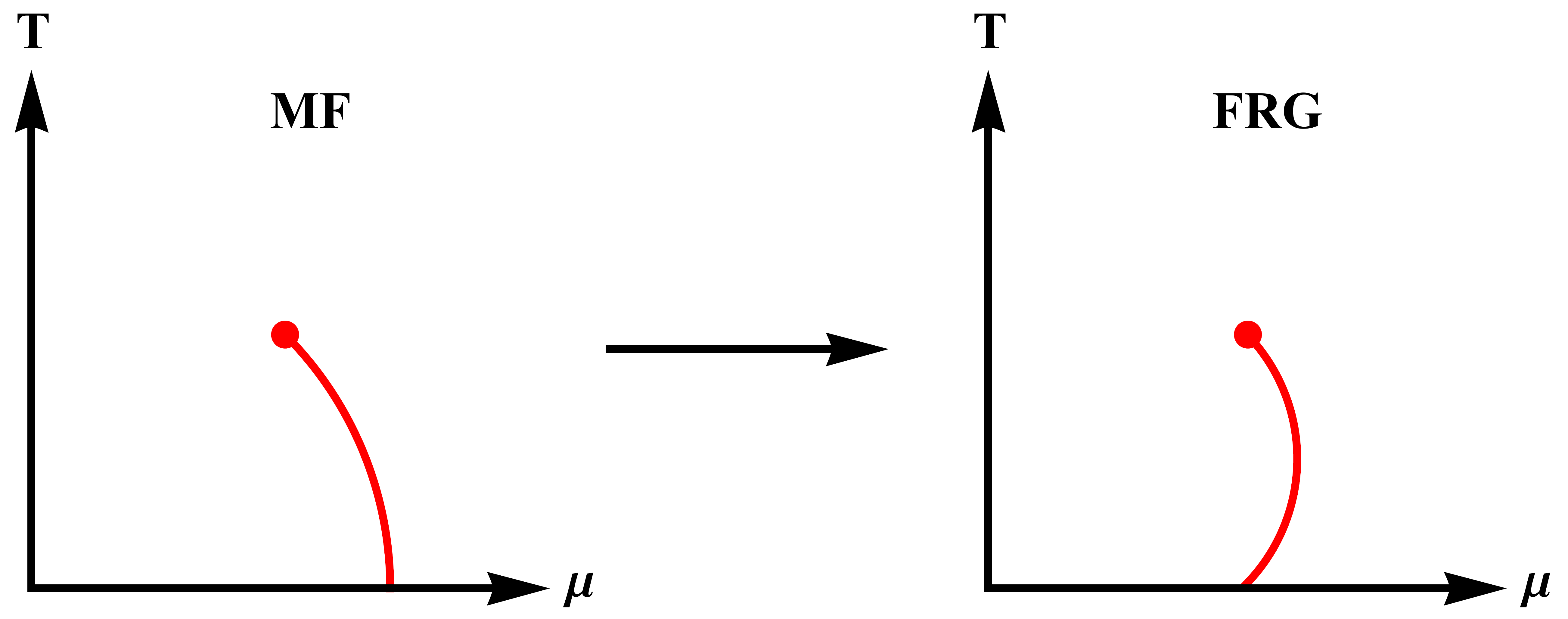}
	\caption{Sketch of the QCD phase diagram in the QM model which features a line of first-order chiral transitions that end in a critical point (CEP) of second order. In mean-field calculations the first-order line exhibits the shape shown on the left-hand side while FRG calculations find the shape on the right-hand side.}
	\label{fig:sketch}
\end{figure}

Standard mean-field calculations in the QM model lead to a first-order phase transition line as sketched on the left-hand side of Fig.~\ref{fig:sketch} where the slope, $dT_c/d\mu_c$, is negative. This is consistent with the Clausius-Clapeyron relation,
\begin{equation}
\label{eq:Clausius_Clapeyron}
\frac{dT_c}{d\mu_c}=-\frac{\Delta n}{\Delta s},
\end{equation}
if one assumes that the changes in quark number density, $\Delta n$, and entropy density, $\Delta s$, are both positive, when going from a dilute gas phase to a liquid-like phase.

The phase diagram of the QM model has also been calculated using the Functional Renormalization Group (FRG) approach which is much more powerful than mean-field theory in describing phase transitions. There it was found that the `Lee-Wick state' of (nearly) massless quarks is avoided in the first-order transition and the sigma field remains finite across the phase boundary~\cite{Schaefer:2004en}. However, at low temperatures, the curvature of the phase boundary bends the `wrong way' in the sense of the Clausius-Clapeyron relation as sketched on the right-hand side of Fig.~\ref{fig:sketch}. For $d\mu_c/dT_c>0$ and $\Delta n>0$ the change in entropy density has to be negative, $\Delta s<0$. 

Although relative to the Stefan-Boltzmann pressure, a decreasing pressure with increasing temperature has been observed in previous FRG studies of the phase structure of the quark-meson model \cite{Schaefer:2004en, Herbst:2013ail, FuPawlowski2015}, this phenomenon, including possible explanations, has not been explored in the literature to our knowledge so far.

Before coming to that, we briefly recall some basics on Eq.~(\ref{eq:Clausius_Clapeyron}). Unlike the usual Clausius-Clapeyron relation for the temperature dependence of the pressure along the transition line, which is proportional to the difference of the entropy per particle on either side of the phase transition, the $\Delta s$ in Eq.~(\ref{eq:Clausius_Clapeyron}) refers to the difference of the entropy per volume, i.e.~the discontinuity of the entropy density. On one hand, in an order-disorder chiral transition one expects the entropy per particle to increase, which together with $\Delta n > 0$ then automatically implies $\Delta s >0$ also, and the distinction between entropy per particle and entropy per volume does not make a qualitative difference. On the other hand, when going from a gaseous to a liquid phase, this distinction becomes important: here the entropy per particle is expected to decrease discontinuously but both signs for $\Delta s$, the entropy-density difference, are possible depending on the size of the density jump $\Delta n$. Only with a sufficient density increase will the entropy density also increase, and hence $\Delta s > 0$, although the entropy per particle decreases.

In that sense, one might conclude that the mean-field result reflects the behavior of a chiral transition while the FRG result is more consistent with a gas-liquid transition to self-bound quark matter, in which $\Delta n$ is simply not large enough, in the quark-meson model. Although this is in-line with the other observations mentioned above, the pathology arises because the size of $\Delta s < 0 $ is larger than the entropy density $s$ on the gaseous side of the first-order line so that the entropy density itself in fact turns negative across this line at low temperatures.    

The same general pattern when comparing mean-field with FRG results was also observed in the parity-doublet model to describe the liquid-gas transition of nuclear matter together with a chiral transition in the high density phase \cite{Weyrich:2015hha}. No thermodynamic inconsistency arises there, however, because the discontinuity in the density at the nuclear-matter transition is always large enough when the saturation density of symmetric nuclear matter is described correctly. 

\section{Theoretical setup}\label{sec:setup}

To set the stage we briefly summarize the pertinent features of the FRG approach for the thermodynamics of the QM model as well as for real-time correlation functions. The latter will be needed to obtain the spectral properties of the in-medium sigma and pion modes which will aid the physical interpretation of the results at low temperatures $T$ and large quark chemical potentials $\mu$. For details of the numerical implementation we refer to our previous work \cite{Tripolt2014, Tripolt2014a}.

The FRG method is based on the Wilsonian idea of integrating out fluctuations associated to different momentum scales $k$. At the ultraviolet (UV) scale $\Lambda$, one has to specify an ansatz for the effective average action $\Gamma_k$, which will be given by the classical action of the QM model. Fluctuations from lower momentum scales are then taken into account by solving the flow equation for the effective action or equivalently the grand potential $\Omega(T,\mu)$ \cite{Wetterich:1992yh, Morris1994},
\begin{equation}
\label{eq:wetterich}
\partial_k \Gamma_k=\frac{1}{2}\text{Tr}\left\{ \partial_k R_k\left(\Gamma^{(2)}_k+R_k\right)^{-1}\right\} .
\end{equation}
The infrared (IR) regulator function  $R_k$ acts as a momentum-dependent mass term and suppresses fluctuations below $k$, $\Gamma^{(2)}_k$ denotes the second functional derivative of the effective average action w.r.t.~the fields, and the trace includes a summation over internal and space-time indices as well as an integration over the loop momenta. In the IR limit $k\rightarrow 0$ the full quantum effective action is obtained.

To make the functional equations tractable, approximations to the effective action are needed. In the present context they amount to a derivative expansion of the effective action, which turns the flow equations into a system of partial differential equations that can be solved numerically. Usually only the lowest-order terms in the derivative expansion are kept, which is referred to as the `local potential approximation' (LPA). In many circumstances it has proven to be reasonably accurate in the calculation of critical phenomena.

In the LPA the Euclidean effective action of the QM model ($N_F=2$)  at a given $T$ and $\mu$ reads
\begin{eqnarray} 
\Gamma_{k}&=& 
\int_x \:\Big\{
\bar{\psi} \left({\partial}\!\!\!\slash +
h(\sigma+i\vec{\tau}\cdot\vec{\pi}\gamma_{5}) -\mu \gamma_0 \right)\psi\nonumber \\
&& \qquad\hspace{4mm}
+\frac{1}{2} (\partial_{\mu}\phi)^{2}+U_{k}(\phi^2) - c \sigma 
\Big\}\nonumber\\
&& \int_x\equiv \int_0^{1/T}dx_0\int_V d^3x.
\label{eq:QM}
\end{eqnarray}
Here $\psi$ denotes the up- and down quark fields. The mesonic pion- and sigma fields are combined as $\phi\equiv(\sigma,\vec{\pi})$, $h$ is the Yukawa coupling and $c\sigma$ an explicit chiral symmetry breaking term. In the LPA only the effective potential $U_k$ flows while, e.g., renormalizations of the kinetic energy, which are of higher order in the derivative expansion, are neglected. At the UV scale 
$\Lambda$ the effective potential is taken to be chirally unbroken, i.e.
\begin{equation}
U_\Lambda(\phi^{2}) =
\frac{1}{2}m_\Lambda^{2}\phi^{2} +
\frac{1}{4}\lambda_\Lambda(\phi^{2})^{2}.
\label{eq:pot_UV}
\end{equation}
and the UV parameters are chosen such as to reproduce a physical pion mass and pion decay constant $f_\pi$ in the IR limit $k\to 0$. The IR quark mass $M_\psi$ is essentially determined by the Yukawa coupling $h$.

The flow equation for $U_k$  can be solved numerically using a grid method (see \cite{Schaefer:2004en} for details) and the grand potential is obtained as $U_{k\to 0}\equiv\Omega (T,\mu)$. From it various thermodynamic quantities such as pressure, entropy density and quark number density can be derived in the standard way:
\begin{align}
p(T,\mu)&=-\Omega (T,\mu)+\Omega (0,0),\\
s(T,\mu)&=\frac{\partial p(T,\mu)}{\partial T},\label{eq:entropy}\\
n(T,\mu)&=\frac{\partial p(T,\mu)}{\partial \mu}.
\end{align}

\begin{figure*}[t]
	\includegraphics[width=0.99\textwidth]{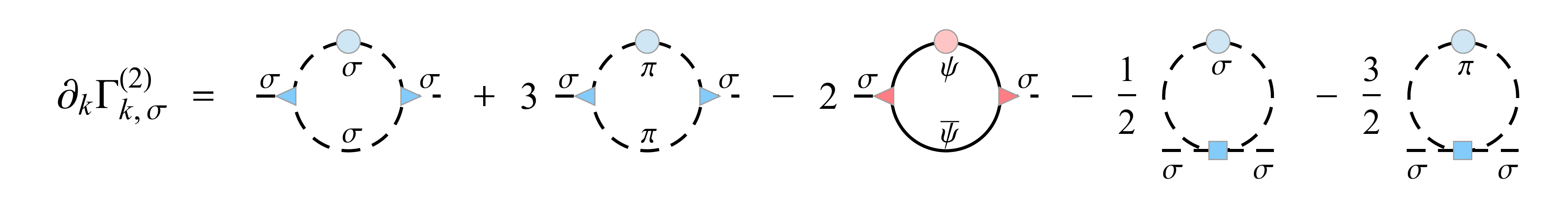}\\
	\includegraphics[width=0.99\textwidth]{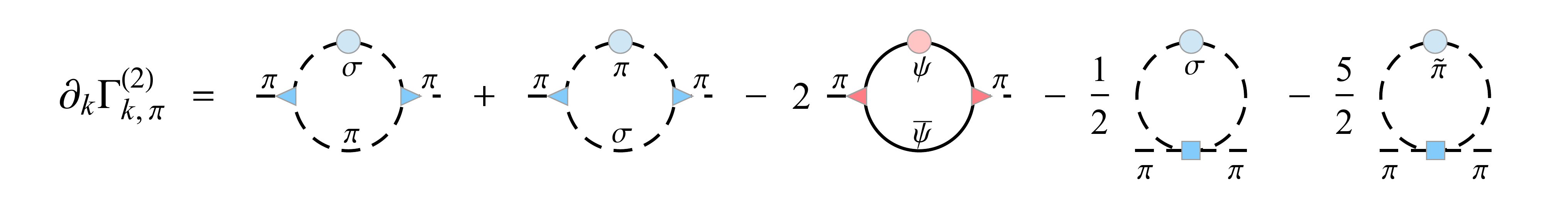}
	\caption{Two-point flow equations for the sigma- and pion-channel of the quark-meson model. Dashed (solid) lines are mesonic (quark) propagators, regulator insertions, $\partial_k R_k$, are labeled with circles while triangles and squares denote three- and four-point vertices, respectively. Figure taken from \cite{TripoltSmekalWambach2017}.}
	\label{fig:flow_equations_gamma2} 
\end{figure*}

In addition to $\Omega (T,\mu)$ we will also need the real-time mesonic spectral functions $\rho_{\pi,\sigma}(\omega,\vec p)$. These are obtained from the Euclidean 2-point functions by the method proposed in \cite{Tripolt2014, Tripolt2014a} which entails an analytic continuation from imaginary to real energies. The flow equations for the 2-point functions $\Gamma^{(2)}_k$ are obtained by taking two functional derivatives of the Wetterich equation, Eq.~(\ref{eq:wetterich}), w.r.t.~the fields as displayed in Fig.~\ref{fig:flow_equations_gamma2}.

As initial conditions for the retarded 2-point functions at the UV scale, we use
\begin{eqnarray}
\label{eq:UV_sigma} 
\Gamma^{(2),R}_{\Lambda,\sigma}(\omega, \vec{p})&=&(\omega+i\epsilon)^2-\vec{p}^{\,2}-2U_{\Lambda}'-4\phi^2 U_{\Lambda}'', \\
\label{eq:UV_pion} 
\Gamma^{(2),R}_{\Lambda,\pi}(\omega, \vec{p})&=&(\omega+i\epsilon)^2-\vec{p}^{\,2}-2U_{\Lambda}',
\end{eqnarray}
where the limit $\epsilon\rightarrow 0$ is assumed implicitly and primes denote derivatives w.r.t.~the field variable $\phi^2$.

The flow equations for the 2-point functions can then be analytically continued to real energies by the
following 2-step procedure. First, the periodicity of the bosonic and fermionic occupation numbers with respect to the discrete external (mesonic) Euclidean energy $p_0=2 n \pi T$ is exploited,
\begin{equation}
\label{eq:continuation1}
n_{B,F}(E+\I p_0)\rightarrow n_{B,F}(E).
\end{equation}
Then, the Euclidean energy $p_0$ is replaced by a continuous real frequency $\omega$,
\begin{equation}
\label{eq:continuation2}
\Gamma^{(2),R}(\omega,\vec p)=-\lim_{\epsilon\to 0} \Gamma^{(2),E}(p_0=-\I(\omega+\I\epsilon), \vec p).
\end{equation}
The spectral functions are then obtained from the retarded 2-point functions as
\begin{equation}
\label{eq:spectralfunction}
\rho(\omega,\vec p)=\frac{1}{\pi}\frac{\text{Im}\,\Gamma^{(2),R}(\omega,\vec p)}{\left(\text{Re}\,
	\Gamma^{(2),R}(\omega,\vec p)\right)^2+\left(\text{Im}\,\Gamma^{(2),R}(\omega,\vec p)\right)^2}.
\end{equation}
For further details on the theoretical setup and the numerical implementation we refer to \cite{Tripolt2014, Tripolt2014a}.

\section{Results}\label{sec:results}

For the results presented below the parameters listed in Tab.~1 have been used, see also \cite{Tripolt2014,Tripolt2014a}. In the present case the constituent quark mass is higher so as to shift the CEP to higher temperatures and to induce a change of curvature of the critical first-order line. 
\begin{table}[h]
	\centering
	\begin{tabular}{C{1.1cm}|C{1.3cm}|C{1.1cm}|C{1.3cm}|C{1.0cm}}
		$\Lambda$ & $m_\Lambda/\Lambda$ & $\lambda_\Lambda$ & $c/\Lambda^3$ & $h$ \\
		\hline
		1 GeV & 0.969 & 0.001 & 0.00175 & 4.2
	\end{tabular}
	\caption{Parameter set used for the QM model used in this work. In the vacuum the following values for the curvature masses and the pion decay constant are obtained in the IR: $\sigma_0\equiv f_\pi=92.5$~MeV, $m_\pi=138$~MeV, $m_\sigma=606$~MeV, $m_\psi=388$~MeV.}
	\label{tab:parameters} 
\end{table}

\begin{figure}[h]
	\includegraphics[width=\columnwidth]{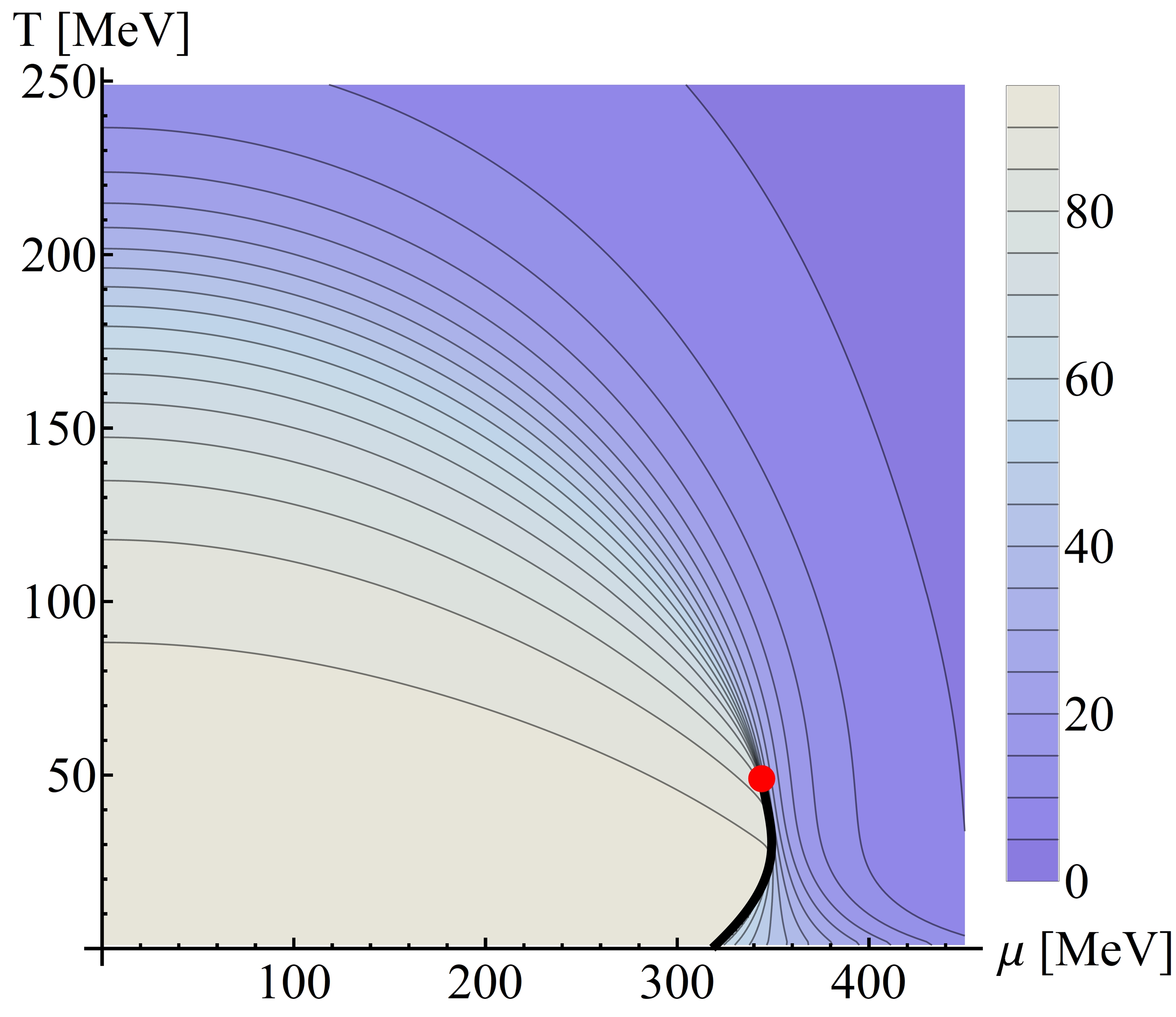}
	\caption{(color online) The phase diagram of the QM model shown as a contour plot of the chiral order parameter $\langle\sigma(\mu,T)\rangle$. The CEP, which is located at $T\approx 50$~MeV and $\mu\approx 345$~MeV, is indicated by the red dot and the first-order transition line is denoted by the solid black line.}
	\label{fig:phase_diagram}
\end{figure}

The phase diagram for the chiral order parameter $\langle\sigma(\mu,T)\rangle$ is shown in Fig.~\ref{fig:phase_diagram}. The first-order line, $T_c(\mu_c)$, has regions with negative and positive slope, which will be of importance in the later discussion. It is also noteworthy that there appears a small triangle-shaped region beyond the first-order line where the $\sigma$ condensate seems to be enhanced. This structure is also present in the chiral limit where it is enclosed between phase transitions of first and second order, as noticed already in \cite{Schaefer:2004en}. 

\begin{figure}[h]
	\includegraphics[width=0.99\columnwidth]{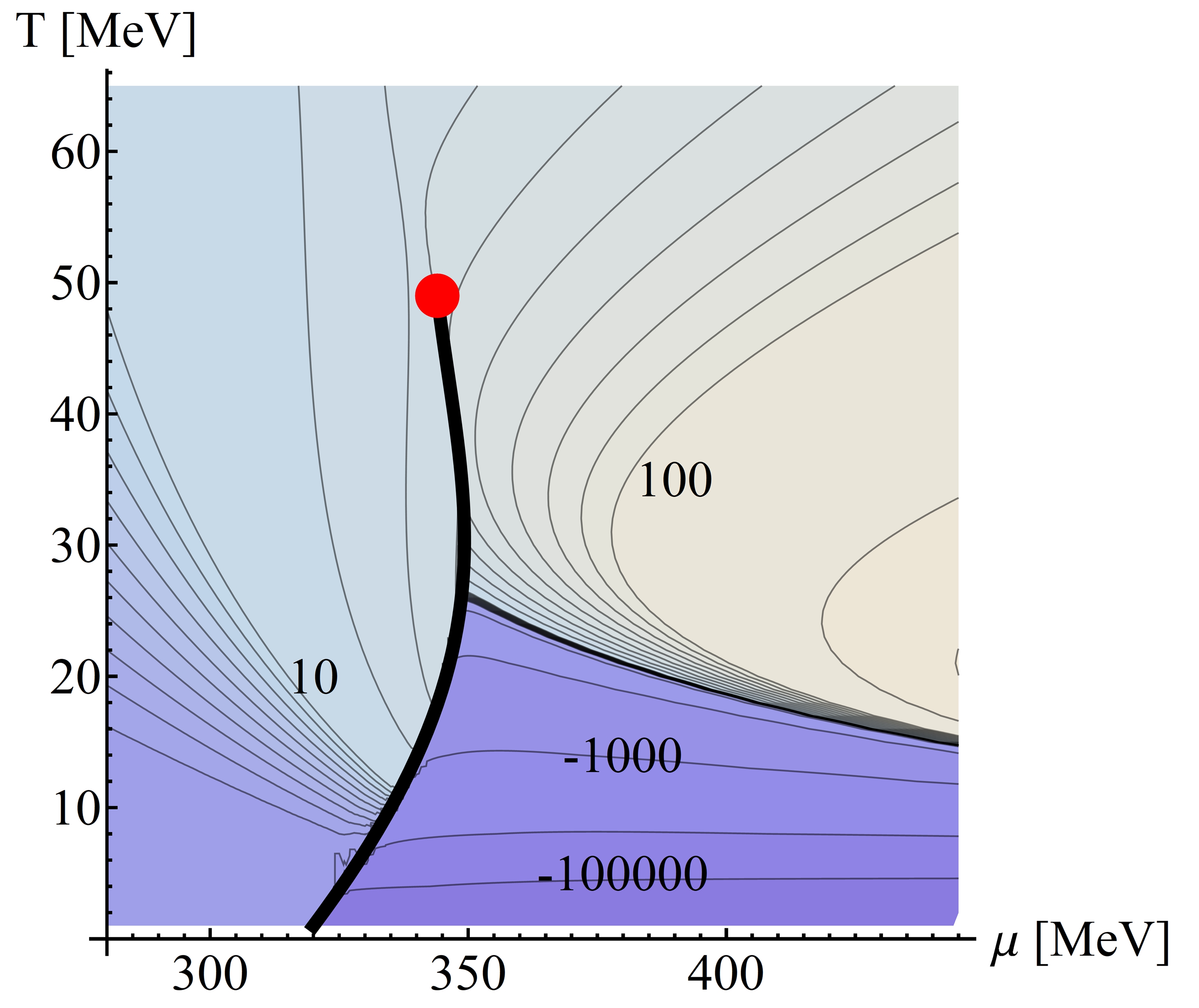}
	\caption{(color online) The dimensionless entropy density, $s/T^3$, is shown as a contour plot together with the CEP and the first-order phase transition line, cf.~Fig.~\ref{fig:phase_diagram}. The inset numbers denote the corresponding values of $s/T^3$. Beyond the first-order line, a regime with negative entropy density is found.}
	\label{fig:entropy_density}
\end{figure}

We note that the positive slope of the first-order line at low temperatures necessitates a decrease of the entropy density as dictated by the Clausius-Clapeyron relation in Eq.~(\ref{eq:Clausius_Clapeyron}).  This in itself is not necessarily worrisome as discussed in the introduction, and there are physical systems which exhibit such a behavior \cite{Iosilevskiy:2015sia}. However, as shown in Fig.~\ref{fig:entropy_density}, beyond the first order phase transition we find an extended region where $s$ is strongly negative. On the critical line this region ends just below the point where the slope $dT_c/d\mu_c$ changes sign (i.e.~where $\Delta s =0$), and it decreases slowly in temperature as $\mu$ increases. By varying the parameters of the QM model and using different 3d-regulator functions we convinced ourselves that the appearance of a $s<0$ region is generic and that its extent is uniquely connected to the slope of the first-order line. By calculating the entropy densities of quarks, sigma mesons and pions separately, we find that the large negative values stem from the quark contribution, while the entropy density of the mesons is always positive. In fact, the entropy density of the sigma mesons is increased in the same region where the quark contribution is negative. 

\section{Discussion}\label{sec:discussion}

Regions of negative entropy density in the phase diagram are thermodynamically unacceptable and possible reasons for this behavior have to be thought of. The first that comes to mind is that this is an artifact of the choice of regulator function \cite{BraunLeonhardtPospiech2017} and/or of the truncation in the derivative expansion of the effective action to lowest order. This is clearly a possibility and has to be studied in the future by going to higher orders. An example for the influence of the truncation on the phase structure can be found in \cite{Zhang:2017icm}. What is more appealing, however, is that there are physical reasons for this problem. Several come to mind and we will discuss them in turn.

\subsection{Color superconductivity}

From extensive studies of the Nambu Jona-Lasinio (NJL) model \cite{Buballa2005} it is known that the Fermi surface of two-flavor quark matter at low temperatures can undergo a pairing transition to a (color)superconducting state. As the QM model in its physical content is very similar to the NJL model this is also expected to happen in the present case. 

To elucidate this point, a schematic model study is added here \cite{Buballa}. When replacing the one-meson exchange diagram by a four-point vertex, where the momentum of the exchanged meson compared to the mass can be neglected, one obtains the following NJL-type Lagrangian
\begin{align}
   \mathcal{L}=\bar{\psi}(i\slashed\partial-m_q)\psi+G_\sigma(\bar\psi \psi)^2+G_\pi(\bar\psi i \gamma_5 \vec{\tau}\psi)^2,
\end{align}
with $G_\alpha=g_\alpha^2/(2m_\alpha^2)$ for $\alpha\in \{\sigma,\pi\}$.  Here we assume that the dynamical quark mass of the order of $m_q=300$~MeV has already been dynamically generated and the pion mass is fixed to $m_\pi=140$~MeV. The coupling constant $g_\pi$ is given by $g_\pi=m_q/f_\pi$ with the pion decay constant $f_\pi=93$~MeV. An explicit value for $g_\sigma$ is not needed in the following. Applying a Fierz transformation to the particle-particle channel along the lines of Ref.~\cite{Buballa2005}, an attractive standard 2SC channel,                         $\langle\psi^TCi\gamma_5\tau_2\lambda_2\psi\rangle$, corresponding to $J^\pi=0^+$, isospin singlet, color anti-triplet, is found. In this channel the coupling constant
\begin{align}
		H_s=\frac{1}{16}(3G_\pi+G_\sigma)=\frac{1}{32}\left(3\frac{g_\pi^2}{m_\pi^2}+\frac{g_\sigma^2}{m_\sigma^2}\right)\ .
\end{align}
is largest and leads to Cooper pairing.
If the contribution of the sigma meson is omitted and the pion mass is kept fixed to its vacuum value this yields a coupling $H_s\approx1/140^2$~MeV$^{-2}$. For a typical NJL UV cutoff of the order of $\Lambda = 600$ MeV the coupling can be estimated by $H_s\Lambda^2\approx 16$ which is considerably larger than typical NJL values of the order of $H_s\Lambda^2\approx 2$, cf.~Ref.~\cite{Buballa2005}.
The conclusion is that there is some evidence for significant superconducting gaps in the quark-meson model. We therefore expect the occurrence of 2SC pairing as soon as a Fermi sphere of quarks has formed, in particular in the low-temperature and high quark-chemical potential regime beyond the phase transition where the anomalous negative entropy density is observed. A more elaborate study of the two-flavor color superconductivity within the quark-meson model based on Eliashberg-type equations for the complex and frequency-dependent pairing gaps will be presented elsewhere \cite{SedrakianTripoltWambach2017}.

To properly treat pairing phenomena within the FRG the fermion propagator, from the start, has to be of the Nambu-Gorkov form, including anomalous off-diagonal terms. Their fluctuations then need to be included within the FRG in addition to those of the chiral order parameter. To our knowledge such flow equations in higher-dimensional field spaces have so far been used in the quark-meson model only for finite isospin density with pion condensation \cite{Kamikado2013} or in the quark-meson-diquark model to describe the competition between chiral and diquark condensate in two-color QCD \cite{Strodthoff:2011tz}.

\subsection{Inhomogeneous phases}

A tacit assumption in the FRG studies of the phase diagram presented above is spatial homogeneity of all phases.  This may be too restrictive. Indeed, from NJL and QM model studies in the mean-field approximation \cite{BuballaCarignano2015,CarignanoBuballaSchaefer2014} it is known that crystalline phases with inhomogeneous chiral order parameters are favored at low $T$ and large $\mu$ over homogeneous ones. A stability analysis of the homogeneous phase should exhibit this fact. Instabilities are signaled by zero-frequency poles of the pertinent retarded real-time propagators at finite spatial critical momenta $\vec p_c$. 

As an example we have looked at instabilities in the `pion-direction', signaled by the inverse static pion propagator  
\begin{equation}
D_\pi^{-1}(\omega=0,|\vec{p_c}|)=0.
\end{equation}
This is the famous example of the onset of `p-wave pion condensation', studied extensively in nuclear matter \cite{EricsonWeise1988a}.  To identify such critical momenta in the FRG we have considered the pion 2-point function (the inverse of $D_\pi$) to search for zeros at intermediate RG scales $k$, i.e.~for 
\begin{equation}
\Gamma^{(2)}_{k,\pi}(\omega=0,|\vec{p_c}|)=0.
\end{equation}

\begin{figure}[b]
	\vspace{.2cm}
	\includegraphics[width=0.48\textwidth]{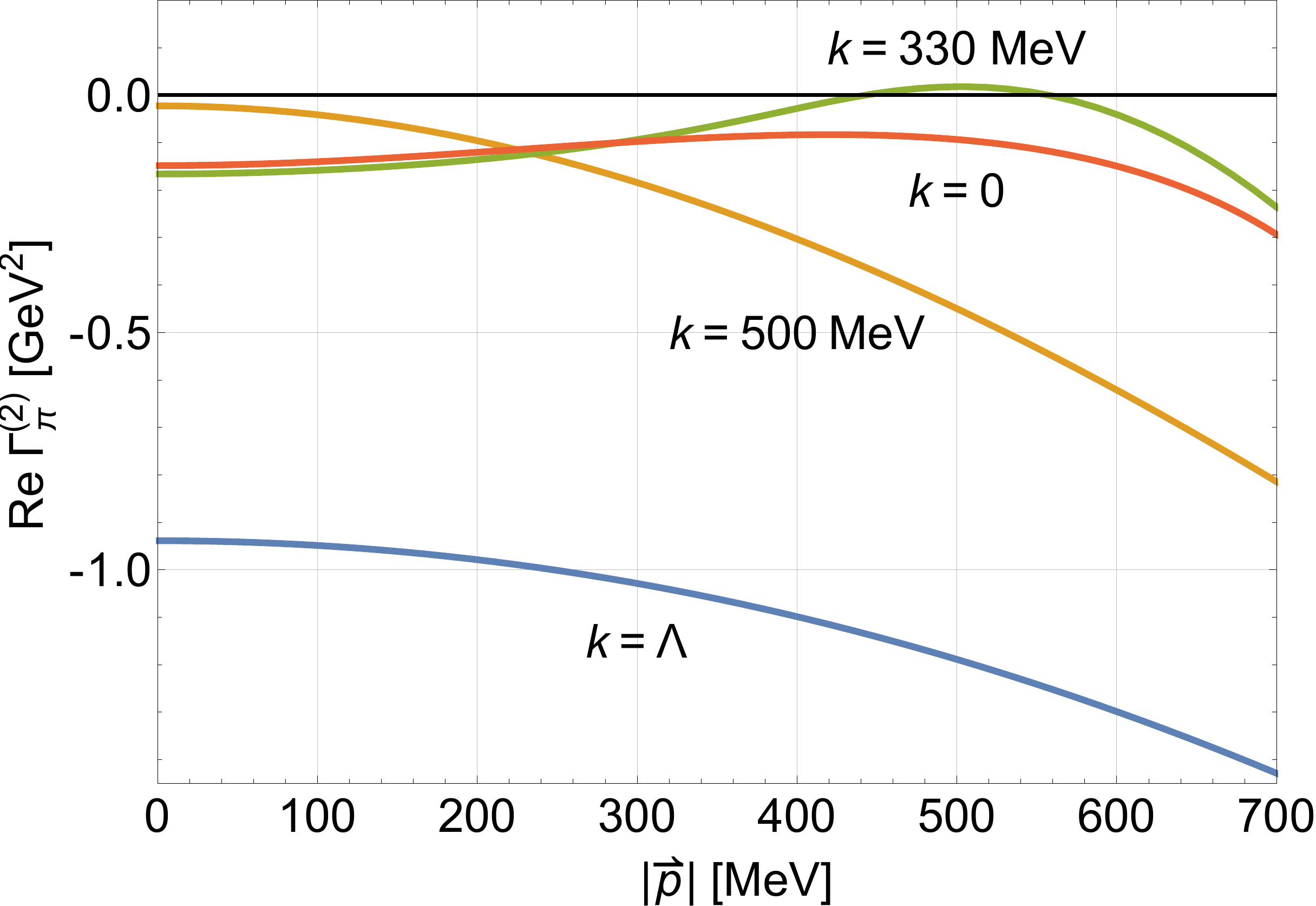}
	\caption{(color online) The real part of the static pion 2-point function, $\Gamma^{(2)}_{k,\pi}(\omega=0,|\vec{p}|)$, evaluated at $\sigma_{0, k_0}$ (see also Fig.~\ref{fig:scale_zero_crossing} and the text for details), is shown for $\mu=400$~MeV and $T=15$~MeV at different RG-scales k: $k=\Lambda=1$~GeV, $k=500$~MeV, $330$~MeV  and $k=0$. A zero at finite $|\vec p_c|$ for intermediate RG scales $k$ is interpreted as evidence of an instability towards formation of an inhomogeneous phase with modulation characterized by $|\vec p_c|$.}
	\label{fig:Gamma2_flow}
\end{figure}

\begin{figure}[h]
	\includegraphics[width=0.47\textwidth]{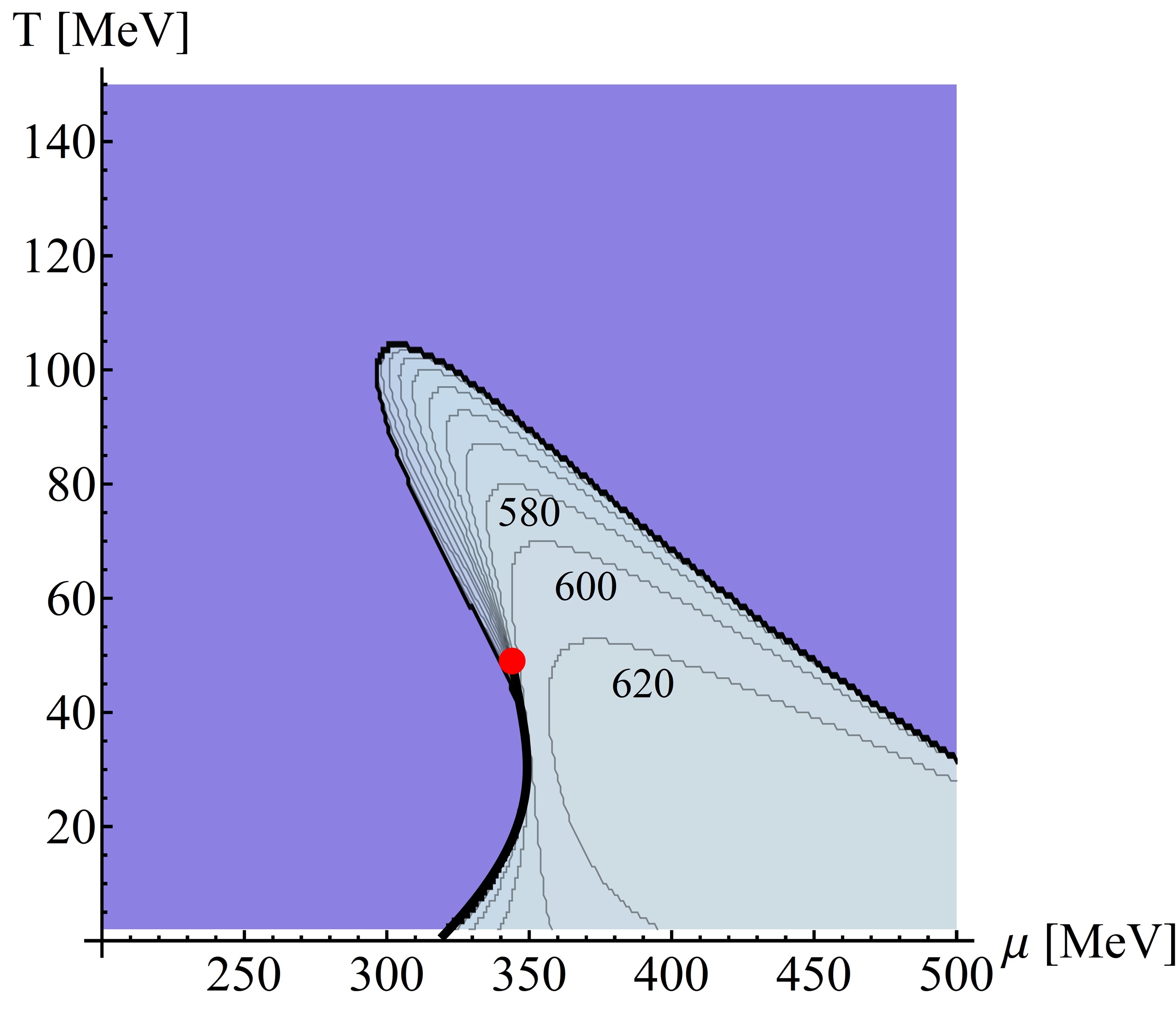}
	\caption{(color online) Regions in the $(T,\mu)$-plane where the real part of the static pion 2-point function, ${\Gamma^{(2)}_{k,\pi}(\omega=0,|\vec{p}|)}$, develops a zero-crossing. The numbers denote the flow scale  $k_0$ where the first zero-crossing in the 2-point function evaluated at $\sigma_{0,\text{IR}}$ appears. For comparison the CEP and first-order transition line is also shown. See text for details.}
	\label{fig:scale_zero_crossing_2}
\end{figure}

\begin{figure}[h]
	\includegraphics[width=0.47\textwidth]{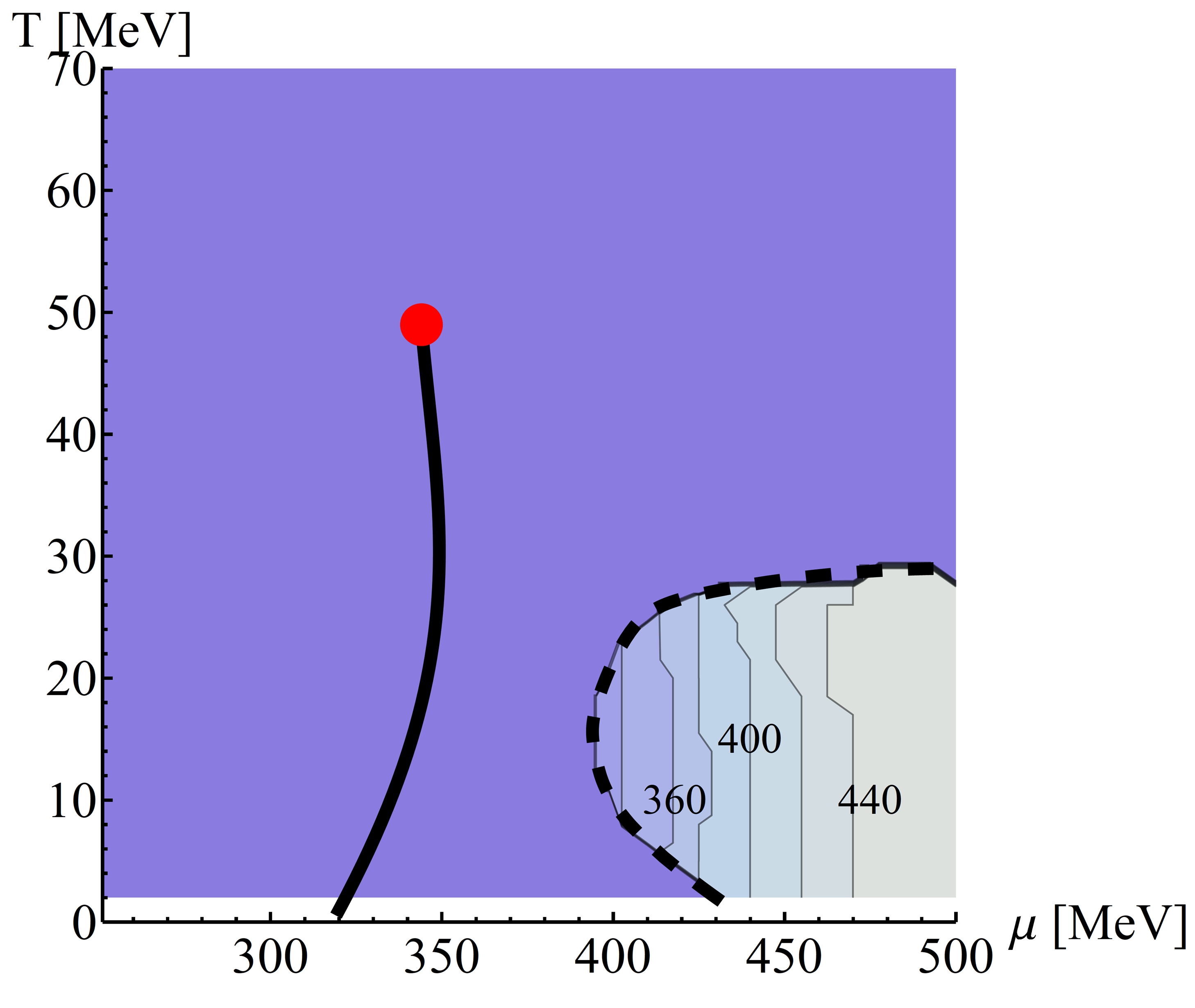}
	\caption{(color online) Regions with zero-crossings in the static pion 2-point function as in Fig.~\ref{fig:scale_zero_crossing_2} but here evaluated at the scale-dependent minima $\sigma_{0, k_0}$ where they first occur.
		The numbers in the dashed region denote the corresponding flow scales $k_0$. See text for details.}
	\label{fig:scale_zero_crossing}
\end{figure}

Thus, we have monitored the FRG flow of the pion 2-point function at finite spatial momenta obtained with the techniques described in Ref.~\cite{Tripolt2014a}. As an example, Fig.~\ref{fig:Gamma2_flow} shows the real part of $\Gamma^{(2)}_{k,\pi}(0,|\vec{p}|)$ for $\mu=400$~MeV and $T=15$~MeV. One observes that a zero-crossing appears at an intermediate RG scale $k$ somewhat above 330~MeV. There the flow equations would probably start to drive the system into this new phase which is however not reachable if the constraint of homogeneity is imposed from the outside. The zero crossing may again disappear in the IR, but this result is meaningless. An analogous behavior was recently also observed as an indication of the formation of inhomogeneous phases in the propagator for collective bosonic fluctuations in the FRG description of imbalanced Fermi gases at unitarity in Ref.~\cite{Roscher:2015xha}.

Based on this insight, we map the region of instability as signaled by a zero crossing in the flow of the static inverse pion propagator $\Gamma^{(2)}_{k,\pi}(\omega=0,|\vec{p}|)$ in the $(T,\mu)$-plane in two different ways. In the first one, we solve the flow equation for the effective potential at a given  $(T,\mu)$-point and identify its minimum at $\sigma_{0,\text{IR}}$ in the IR.  Afterwards we evaluate the flow equation for the pion 2-point function $\Gamma^{(2)}_{k,\pi}(0,|\vec{p}|)$ on this minimum. The region where zero crossings for some momentum $p = p_c$ occur and the corresponding scales $k_0$ at which they occur inside this region are plotted in Fig.~\ref{fig:scale_zero_crossing_2}. We note that the corresponding region for the sigma 2-point function is basically the same except that it is a bit smaller. It only extends to about the $k = 580$~MeV line in Fig.~6. Beyond this line $p$-wave pion condensation therefore seems to be favored over sigma condensation, which is likely due to the residual mass difference between the two in this region. The masses of sigma and pion approach one another only gradually with increasing temperature and chemical potential as chiral symmetry gets effectively restored.

The second way to visualize the instability in the $(T,\mu)$-plane relies on a temporary stop of the potential evolution at some intermediate scale $k > k_{\text{IR}}$ with the identification of the corresponding potential minimum. Subsequently, the evolution of the 2-point functions evaluated for this minimum are also stopped at this intermediate scale. This procedure is repeated for various scales $k$ towards the IR. The region of zero crossings for some momenta and the scales at which they occur in this way are recorded in Fig.~\ref{fig:scale_zero_crossing}.
This procedure has some similarity with a co-moving Taylor-expansion solution wherein the potential flow is evaluated at a scale-dependent minimum.
However, as expected both procedures agree in the infrared.

The regimes in Figs.~\ref{fig:scale_zero_crossing_2} and \ref{fig:scale_zero_crossing} provide evidence of the appearance of a spatially modulated ground state such that within these regions in the phase diagram inhomogeneous phases might be favored over homogeneous ones. The scales $k_0$ at the onset thereby serve as rough estimates of the scale of the modulation that might arise in the ground state, i.e.~the wave number of a chiral spiral or a chiral density wave for example. The two different ways to estimate such instabilities towards a spatially modulated phase give a rough feeling for the uncertainty in the translation of an intermediate RG scale $k_0$, without precise physical meaning, into a scale of modulation in the ground state.

In order to map the region of such instabilities with fluctuations beyond mean field from the FRG more precisely, a truncation that allows for a spatially modulated chiral order parameter will be needed in the future.

\section{Summary}\label{sec:summary}

In FRG studies of the QCD phase diagram based on the QM model a region of negative entropy density is found at small $T$ and large $\mu$. It is bounded by a first-order transition line of positive curvature, a feature that is absent in the mean-field approximation and has first been found in Ref.~\cite{Schaefer:2004en}. The region of negative entropy density which is robust against choices of parameters and 3d-regulator functions clearly points to problems with `traditional' treatments the QM model phase diagram within the FRG. Several reasons come to mind. First, the lowest-order truncation in the derivative expansion of the effective action might be insufficient and higher-order could remedy the problem. This possibility needs to be addressed in the future. A more intriguing alternative is of physical nature in that the basic physical assumptions that go into the effective action are not fulfilled and the FRG correctly identifies this through the wrong thermodynamics. Since the negative entropy is caused in the fermion sector an obvious possibility is a Cooper instability of the Fermi surface from attractive pion and sigma exchange. Indeed it could be verified that the region of negative entropy density features (color) superconducting pairing gaps. To properly account for pairing the fermion part of the effective action has to contain Nambu-Gorkov propagators and additional flow equations for the corresponding diquark channel have to be solved.
As anticipated from QM model studies of spatial inhomogeneities of the chiral order parameter in the mean-field approximation there is also the possibility that the tacit assumption of spatially homogeneous phases is incorrect. This would entail FRG calculations with spatial inhomogeneities much like as in condensed matter physics. These are clearly much more challenging numerically. Comparing Figs.~\ref{fig:entropy_density} and \ref{fig:scale_zero_crossing} it seems even quite plausible that a combination of both, a pairing instability and inhomogeneities might be possible in different regimes thus leading to a rather rich phase diagram already in the quark-meson model.  

To conclude, we have shown that the 'standard' treatment of chiral effective models of QCD, such as the QM model has serious problems at low temperatures and large chemical potentials within the FRG. Most likely several physical effects have to be taken into account. In addition to the possibilities discussed in this note, it is obvious that at low temperatures nucleonic degrees of freedom are of utmost importance which complicate the problem even further.

\acknowledgments 
We thank Michael Buballa for useful discussions and for providing the estimates for the gap of the superconducting CSC phase. We also
thank Armen Sedrakian for preliminary computations of the pairing gaps at high chemical potentials and small temperatures based on Eliashberg-type equations.
This work was supported by the BMBF grants 05P15RGFCA and 05P16RDFC1, and by HIC
for FAIR within the LOEWE program of the State of Hesse.
B.-J.~S. acknowledges support by the FWF grant P24780-N27. L.~v.~S. and J.~W. acknowledge support by the Deutsche Forschungsgemeinschaft (DFG) through the CRC-TR 211 ``Strong-interaction matter under extreme conditions''.

\bibliography{negative_entropy_final_resubmit}

\end{document}